\documentclass[10pt,letterpaper,compsoc,conference]{iiswc24}

\usepackage{cite}
\usepackage{amsmath,amssymb,amsfonts}
\usepackage{algorithmic}
\usepackage{graphicx}
\usepackage[dvipsnames]{xcolor}
\usepackage[final]{microtype}
\usepackage[italic]{mathastext}
\usepackage{libertine}
\usepackage[T1]{fontenc}
\usepackage{textcomp}
\usepackage[varqu,varl]{zi4}
\usepackage[all]{nowidow}
\usepackage[keeplastbox]{flushend}
\usepackage{fancyhdr}

\usepackage{authblk}
\usepackage{listings}
\usepackage{xcolor}
\usepackage{lipsum}
\usepackage[most]{tcolorbox}

\definecolor{codegreen}{rgb}{0,0.6,0}
\definecolor{codegray}{rgb}{0.5,0.5,0.5}
\definecolor{codepurple}{rgb}{0.58,0,0.82}
\definecolor{backcolour}{rgb}{0.95,0.95,0.92}
\lstdefinestyle{mystyle}{
    backgroundcolor=\color{backcolour},   
    commentstyle=\color{codegreen},
    keywordstyle=\color{magenta},
    numberstyle=\tiny\color{codegray},
    stringstyle=\color{codepurple},
    basicstyle=\ttfamily\footnotesize,
    breakatwhitespace=false,         
    breaklines=true,                 
    captionpos=b,                    
    keepspaces=true,                 
    numbers=left,                    
    numbersep=5pt,                  
    showspaces=false,                
    showstringspaces=false,
    showtabs=false,                  
    tabsize=2
}

\usepackage{xcolor}
 
\fancypagestyle{firstpage}{
  \fancyhf{}
  
  \fancyfoot[C]{\thepage}
}

\begin{document}

%% EDIT TITLE BELOW

%\title{Hardware-Agnostic Forecasting of LLM Inference Efficiency via Analytical Modeling}
\title{Forecasting LLM Inference Performance via Hardware-Agnostic Analytical Modeling}

%% DO NOT EDIT THE FOLLOWING

%\renewcommand\Authsep{\qquad}
%\renewcommand\Authand{\qquad}
%\renewcommand\Authands{\qquad}

%% EDIT AUTHOR LIST BELOW

%\author{ }
\author{Rajeev Patwari}
\author{Ashish Sirasao}
\author{Devleena Das}
\affil{Advanced Micro Devices (AMD), San Jose, California}

%%% ALTERNATIVE FORMAT FOR MULTIPLE SCHOOLS:
%%% 
% \author[1]{Author1 Name}
% \author[2]{Author2 Name}
% \author[2]{Author3 Name}  
% \author[1]{Author4 Name}
% \affil[1]{Full Name of Awesome School}
% \affil[2]{Full Name of Awesomer School}

\maketitle
\thispagestyle{firstpage}
\pagestyle{plain}

%% EDIT YOUR PAPER'S CONTENTS BELOW

\begin{abstract}
Large language models (LLMs) have been increasingly deployed as local agents on personal devices with CPUs, NPUs and integrated GPUs. However, forecasting inference performance on devices with such heterogeneity remains challenging due to the dynamic compute and memory demands. Existing approaches rely on GPU benchmarking or machine learning-based latency predictors, which are often hardware-specific and lack generalizability. To this end, we introduce LIFE, a lightweight and modular analytical framework that is comprised of modular analytical model of operators, configurable to characterize LLM inference workloads in a hardware and dataset-agnostic manner. LIFE characterizes the influence of software and model optimizations, such as quantization, KV cache compression, LoRA adapters, chunked prefill, different attentions, and operator fusion, on performance metrics such as time-to-first-otken (TTFT), time-per-output-token (TPOT) and tokens-per-second (TPS). LIFE enables performance forecasting using only hardware specifications, such as TOPS and memory bandwidth, without requiring extensive dataset benchmarking. We validate LIFE's forecasting with inference on AMD Ryzen CPUs, NPUs, iGPUs and NVIDIA V100 GPUs, with Llama2-7B variants, demonstrating the utility of LIFE in forecasting LLM performance through lens of system efficiency to enable efficient LLM deployment across different hardware platforms. 

\end{abstract}

\section{Introduction}
The Generative Pre-Trained Transformer ~\cite{openai2022chatgpt}, a decoder only LLM architecture based on the original Transformer architecture ~\cite{vaswani2017attention}, has become the foundation for modern large language models (LLMs) such as Phi-4 ~\cite{abouelenin2025phi}, Llama2 ~\cite{touvron2023llama} and Llama3 ~\cite{grattafiori2024llama}. Recent advancements from Deepseek-V3 \cite{liu2024deepseek}, have shown that smaller LLMs, with only a few billion parameters, can offer improved performance compared to earlier models of similar size. Furthermore, parameter-efficient fine-tuning (PEFT) ~\cite{pope2023efficiently} techniques, such as Low Rank Adaptation (LoRA) \cite{hu2021loralowrankadaptationlarge} has enabled efficient LLMs for task-specific applications. These efforts have led to an explosion of interest in local LLMs for faster and private on-device inference on laptops and mobile devices with heterogeneous hardware comprising of Neural Processing Units (NPU), Graphics Processing Units (GPU) and Central Processing Units (CPU), like AMD Ryzen APUs ~\cite{amdryzenai}. 

% State of the art training and distillation techniques as published in DeepSeek-V3 ~\cite{liu2024deepseek} show that smaller LLMs with only a few billion parameters offer improved performance compared to predecessors of similar parameter size. Furthermore, the pretrained LLMs can easily be finetuned to specific applications using Parameter Efficient Finetuning (PEFT) ~\cite{pope2023efficiently} techniques such as Low Rank Adaptation (LoRA) \cite{hu2021loralowrankadaptationlarge}. 
% This has resulted in explosion of interest in local LLMs for faster and private on-device inference on laptops and mobile devices with heterogeneous hardware comprising of Neural Processing Unit (NPU), Graphics Processing Unit (GPU) and Central Processing Unit (CPU), like AMD Ryzen APUs ~\cite{amdryzenai}. 

Enabling efficient LLM inference on personal devices with CPUs, NPUs and integrated GPUs remains challenging due to architectural heterogeneity with distributed and limited memory bandwidth, and difficulties in forecasting performance due to the lack of hardware and dataset-agnostic performance models. Modern heterogeneous devices contain distinct memory and hardware specifications of compute capacity in Tera Operations Per Second (TOPS, or TOPs/sec) and bandwidth, Giga Bytes per second (GBps). The peak performance and memory utilization specifications are not sufficient to forecast performance as the efficiency of hardware varies with the inherent dynamism in the LLM inference workload. Specifically, LLM inference workloads are non-uniform where compute and memory demands vary with prompt and generation lengths, and token generation latency change over time due to KV cache growth. For example, TOPs required for prompt length 128 is much less than 2048 tokens. 
Similarly, time per output token (TPOT) of the first token generated is different from that of 1000th token due to increased memory in KV Cache. While software optimizations such as operator fusion and model optimizations such as KV cache compression \cite{hooper2025kvquant10millioncontext}, quantization \cite{lin2024awqactivationawareweightquantization} to 4-bit, micro-scaling precision formats ~\cite{rouhani2023microscaling} and attention mechanisms can boost inference performance, it is imperative to understand their influences on performance prediction. Existing work in the area of LLM performance forecasting focuses on GPUs like NeuSight ~\cite{Lee_2025}, ~\cite{Cho_2024} and ASTRA \cite{won2023astra} that use simulators to get insight into heterogeneous architectures or GPUs. However, there is a gap in understanding dynamic nature of LLM workload on hardware efficiency and vice versa.  This brings us to the core motivation of our work to answer the following: \textit{(1) What are the fundamental workload requirements for LLM inference? (2) What types of dynamically changing conditions emerge in LLM inference? and (3) How do variations in hardware efficiency impact these dynamic behaviors and affect LLM inference performance?}

In this work, we present the \textbf{LLM Inference Forecast Engine (LIFE) Framework}, a lightweight analytical framework for modeling LLM inference workloads in a hardware and data-agnostic manner. LIFE consists of  analytical models of core LLM operators, which are then utilized to build a configurable analytical workload model that supports various datatypes, software and model optimization techniques ~\cite{li2025surveylargelanguagemodel}. Using the analytical LLM workload model, LIFE simulates and quantifies the dynamic compute and memory utilization across the LLM inference timeline. We showcase LIFE's  workload characterization capabilities across variants of Llama2-7b\cite{touvron2023llamaopenefficientfoundation}, focusing on both the LLM decode and prefill phase, and quantify the impact of software and model optimizations and demonstrate the impact of variable hardware efficiency on performance. Our results reveal critical bottlenecks in hardware utilization, highlighting opportunities in designing improved and optimized hardware for LLM inference. The contributions of our work is summarized as follows:
\begin{enumerate}
\item \textbf{Hardware and dataset agnostic analytical framework for LLM inference}: We introduce LIFE, a set of modular operator-level analytical models that can be configurable to support different datatypes, model and software optimizations to capture the dynamic compute and memory behaviors of LLM inference. 
\item \textbf{Characterization of LLM inference phases with optimization trade-offs}: LIFE provides empirical analysis of the prefill and decode phases under varying prompt/generation lengths and optimizations including quantization, KV cache compression, chunked prefill, different attention mechanisms and operator efficiencies. 

\item \textbf{Forecasting hardware-aware LLM performance without datasets}:
LIFE predicts TTFT, TPOT and TPS using only TOPS and bandwidth, enabling hardware-aware performance estimation without requiring benchmarking datasets. We very forecasted performance with true inference of Llama2-7B on AMD Ryzen CPU, NPU, iGPU and NVIDIA V100 GPU hardware.

\end{enumerate}

\section{Background}
% LLM inference optimization is evolving to address (1)  inherent dynamic characteristics of prefill and decode stage in LLMs, (2) faster inference of parameter-efficient finetuned models ~\cite{hu2021loralowrankadaptationlarge},
% %finetuned LLMs that add more parameters to base model 
% (3) and increasing model sizes. There is a need to characterize the influence of these factors on workload efficiency and our LIFE framework enables these analyses in a hardware and dataset independent manner. Below we first discuss the importance of these dimensions for understanding LLM workload efficiency.
There are several factors that can influence workload efficiency which we motivate below. Our LIFE framework enables these factors and provides workload analysis and forecasting in a hardware and dataset independent manner. 

\subsection{LLM Architecture}
\begin{figure}[htbp]
\centerline{\includegraphics[width=\columnwidth]{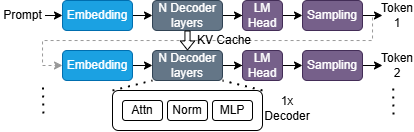}}
\caption{LLM Architecture}
\label{llm-workload}
\end{figure}

% A simplistic view of LLM architecture is shown in Fig.~\ref{llm-workload}. 
Fig.~\ref{llm-workload} shows a simplified view of the LLM architecture, highlighting the two distinct phases in inference: prefill and decode phase. In the prefill phase, the input is processed to set the context for the model in KV Cache ~\cite{pope2023efficiently}. During decode, a new token is generated auto-regressively, updating the KV cache.
% An LLM is built by chaining multiple decoders sequentially. 
% LLMs are causal and operate in two distinct phases (1) prefill phase, during which the input is processed to set the context for the model in KV Cache ~\cite{pope2023efficiently} and (2) decode phase during which a new token is generated auto-regressively, updating the KV cache. 
The fundamental building blocks of LLMs are Embedding layer, consecutive Decoder layers, followed by a Language Model (LM) Head layer and sampling, with Elementwise and Normalization layers in between. The Attention, MLP and normalization layers together form the Decoder layer. Attention layer comprises of Query, Key, Value computations, Rotational Position Encoding (RoPE), an attention mechanism, i.e. either Multi Head-Attention (MHA) \cite{vaswani2017attention}, Group-Query Attention (GQA) \cite{ainslie2023gqatraininggeneralizedmultiquery}, Multi Query Attention (MQA) ~\cite{shazeer2019fast} or Multi-Head-Latent-Attention (MLA) ~\cite{liu2024deepseek} followed by output projection. The MLP consists of projection layers and an activation function. These finite operators in LLMs enable us to develop a low footprint analytical model within the LIFE framework.

% \subsection{LLM Inference resource utilization} 
%Additionally, LIFE quantifies the memory and compute utilization during the prefill and decode phases. Prefill performance depends on prompt length, while token generation slows down in the decode phase as the KV cache grows. Understanding the resource utilization enables LIFE to forecast LLM performance from a compute and memory efficiency point-of-view. 

% quantifying the memory and compute utilization during both prefill and decode phases gives us ability to forecast the performance. Prefill phase performance varies with prompt length. During decode phase, as KV cache increases, token generation slows down as more tokens are generated. Using the LIFE framework, we quantify the resource utilization, accumulate them in a statistics database and analyze the results forecast LLM performance through lens of compute and memory efficiency. 

\subsection{Software optimizations for inference}

The following software optimizations are modeled in LIFE. Operator fusion and memory reuse are widely adopted software optimizations for inference speed up. Operator fusion reduces kernel dispatch calls, kernel launch (dispatch) latency and memory utilization between sequential operations in a model. Operator fusion results in reduced memory access, but the compute operations remain unchanged. Flash Attention \cite{dao2022flashattentionfastmemoryefficientexact} is an example of operator fusion for performance improvement.
% that demonstrates the power of operator fusion for performance improvement. 
Approximations of activation functions using polynomial or piecewise linear methods ~\cite{10650022} is another optimization technique used in Single Instruction Multiple Data (SIMD) accelerators.

\subsection{Model optimizations}

\textbf{Model quantization} reduces the total memory footprint of model parameters. Since generating each new token in an LLM requires a full model pass, reducing model size directly improves token generation throughput.
% new token generation in LLMs requires the entire model to be executed once, reduction in model size consequently improves token generation throughput.
GPTQ ~\cite{frantar2022gptq} and AWQ ~\cite{lin2024awqactivationawareweightquantization} quantization algorithms enable weight reduction from FP16/BF16 to 4-bit without significant loss of accuracy.  

\textbf{KV Cache compression} is another model optimization technique that reduces the memory footprint of the KV cache.
% One of the compression techniques is quantizing the KV cache.
Reducing size of KV cache reduces the amount of history read for every new token generation, thus reducing TPOT and increasing the TPS. The KV cache can be compressed to 4-bit or 8-bit ~\cite{hooper2025kvquant10millioncontext}, whereas MLA exhibits another form of KV compression without quantizing the contents of KV cache ~\cite{liu2024deepseek}.

In LIFE, we analyze compute and memory utilized with or without KV cache compression and weight quantization.

\subsection{LoRA Finetuned LLMs}
LoRA \cite{hu2021loralowrankadaptationlarge} is a PEFT method that updates two smaller matrices during finetuning, as opposed to all trainable parameters. Finetuned adapters are merged with a pretrained base model's weights either prior to inference as a single step, or dynamically merged for every single GEMM operator call \cite{lorahf}. In LIFE, we enable analyzing workload and performance impact for both dynamic and static ahead-of-time LoRA adapter merging

% LIFE framework allows us to analyze workload and study impact to performance for both cases, dynamic LoRA adaptation during inference and static ahead of time merging of adapters with base model. 

\section{Methodology}
\begin{figure*}[htbp]
\centerline{\includegraphics[width=\textwidth]{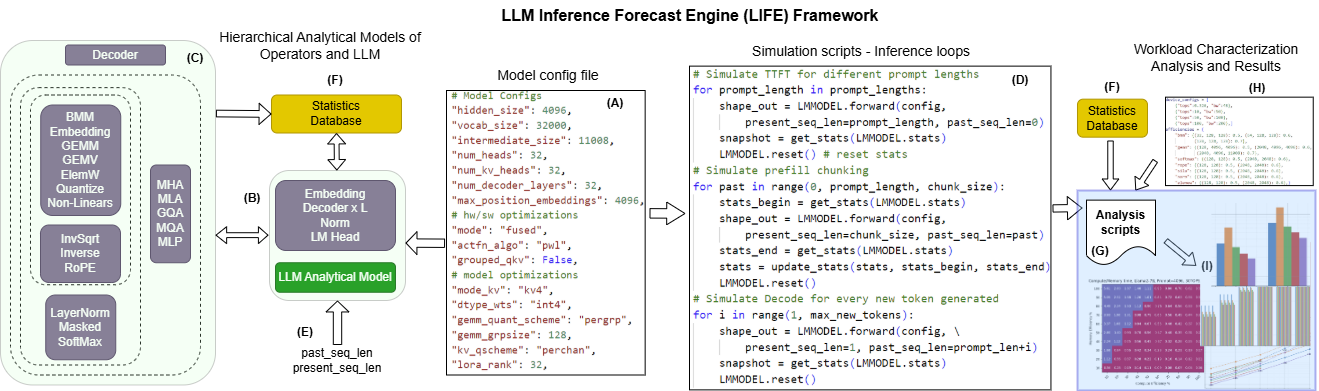}}
\caption{Overview of the LIFE framework. (A) configuration file defines operating conditions and optimizations. (B) Analytical LLM model is built using (C) hierarchical operators. (D) Simulation scripts run LLM scenarios with input sequence lengths (E). (F) The statistics database collects hardware and dataset agnostic metrics and (G) analysis scripts combine these with (H) hardware specs to forecast performance (I).}
\label{life}
% (1) Configurable hierarchical analytical models of operators (2) Statistics database accumulates operator utilization during simulation (3) Configurable analytical model of LLM (4) Operating conditions of history and current sequence lengths (5) Model configuration file to configure the LLM model to operate like Llama2-7B or Phi-4, etc. with optimization strategies like fusion, KV quantization, etc. (6) Simulation scripts orchestrate simulation of analytical model and gather statistics (7) Hardware configuration for forecasting performance (8) Analysis scripts utilize the simulation statistics and forecasts LLM performance for hardware specifications (9) Experimental results of efficiency based TTFT, TPS forecast. 
\end{figure*}

Fig.\ref{life} presents LIFE which takes as input (A) a config file that specifies the operating conditions and optimizations to configure (B) the analytical LLM, comprised of hierarchical analytical models of operators shown in (C). The simulation scripts (D) simulate LLM scenarios defined by the config file and input operating conditions of past/present sequence lengths (E). The analytical LLM and operator models update the statistics database (F) and the analysis script (G) analyzes metrics from the database, with hardware specifications described in (H) to analyze workload characterization and forecast performance metrics (I). The statistics database collect metrics for characterization and forecasting that are hardware agonistic and dataset independent, enabling LIFE's generalizability to different hardware.

\subsection{Analytical Model of Operators}
The lightweight analytical model of operators in LIFE measures compute and memory utilization required by each operator based on input conditions. This abstraction makes LIFE hardware and data agnostic, as actual computations are not computed, allowing to quickly gather information across various LLM operating configurations. Data movement operations like resize and transpose are considered as fused to the compute operations in LIFE's analytical model. We categorize operators into foundational and derived types. The analytical model assumes that if the tensors do not fit into the on chip chip, they are tiled and executed on accelerator which increases the number of dispatch calls. LIFE's framework provides means to model and measure dispatch calls.

\textbf{Foundational Operators} are computed on hardware with a single operation. Table~\ref{foundational_ops} provides a list of foundational operators modeled in LIFE and the corresponding compute and memory utilization, which is a function of datatype, represented by $nbytes$ and $qbytes$ (eg. $nbytes=2$ for bf16 and $qbytes=0.5$ for 4-bit). The input tensor shapes are generally denoted by $(m,k) and (k,n)$ for GEMM, $(m,n)$ for elementwise, $(b,m,k), (b,k,n)$ for BMM, etc.  Non-linear approximations \cite{horner} generally implemented on accelerators are also modeled. The analytical model of the Linear operator is shown as an example in appendix-\ref{appendix-a}.

\begin{table}[ht]
\caption{Foundational Operators}
\centering
\fontsize{10}{10}\selectfont
\begin{tabular}{|p{2cm}|p{1.4cm}|p{3.9cm}|}
\hline
\textbf{Operator} & \textbf{\textit{Compute Ops}}& \textbf{\textit{MemRD + MemWR[Bytes]}} \\
\hline
Linear (GEMM+Bias) & $2mkn$ & $\left(\left(mk \right)+\left(kn\right)+n\right) +\left(mn\right)$\\
\hline
(De)Quantize (Shift+Scale)& $2num\_el$ & $num\_el \times nbytes + num\_qparams \times nbytes + num\_el \times qbytes$ \\
\hline
BMM& $2bmkn - bmn$ & $((bmk)+(bkn)) \times nbytes + (bmn) \times nbytes$ \\
\hline
Elemw & $mn$ & $2mn \times nbytes + mn \times nbytes$ \\
\hline
Non-Linear (Piece Wise Linear) & $2num\_el$ & $\left(num\_el +tables\right)\times nbytes + num\_el \times nbytes$ \\
\hline
Non-Linear (Polyonimal Approx.) & $((n(n+1) /2)+n)$ $\times num\_el$ & $\left(num\_el+n\right) \times nbytes + num\_el \times nbytes$ \\
\hline
Embedding & $1$ & $vocabsize \times $ $hiddensize \times nbytes + hiddensize$ $\times nbytes$ \\
\hline
\end{tabular}

\label{foundational_ops}
\end{table}

\textbf{Derived Operators} in LLM are derived from one or more foundational operators, listed in Table~\ref{derived-ops}. Eg. MHA is a combination of BMM, softmax and element wise add, multiply operations. inverse ~\cite{8525803} and inverse square root ~\cite{fastinvsqrt} approximations are also modeled.

\begin{table}[ht]
\caption{Derived Operators}
\centering
%\renewcommand{\arraystretch}{1.3}
%\resizebox{\columnwidth}{!}{%
\fontsize{10}{10}\selectfont
\begin{tabular}{|p{2.2cm}|p{5.5cm}|}
\hline
\textbf{Operators} & \textbf{\textit{Foundational Operator Used}}\\
\hline
(Quantized) Linear & Linear (GEMM+Bias), Dequantize (Elemw Add, Mul) - int4, int8, int16, MXFP8, MXINT8, etc. \\
\hline
(Quantized) LoRA Linear& Linear (GEMM+Bias), Elemw Add, MatMul with optional LoRA \\
\hline
Inverse Square-root & Elemw Add, Mul \\
\hline
Inverse & Elemw Add, Mul \\
\hline
RoPE & Elemw Add, Mul \\
\hline
Norm & Elemw Add, Mul, Inverse\\
\hline
Softmax & NLF, Elemw Add, Mul, Inverse \\
\hline
MLP & Linear (GEMM+Bias), NLF Elemw Add, Mul \\
\hline
MHA & BMM, Softmax, Elemw Add, Mul \\
\hline
MLA & Linear, ElemW Add, Mul, MHA, Norm, RoPE \\
\hline
\end{tabular} 
\label{derived-ops}
\end{table}

\subsection{Software Optimizations in LIFE}
% In LIFE, we model operator fusion, shape padding, and memory reuse software optimizations to analyze LLM workload. 

\subsubsection{Operator Fusion \& Memory Reuse}
Operator fusion is analytically modeled by removing memory access between operations. For example, with a fused MHA, like Flash Attention \cite{dao2022flashattentionfastmemoryefficientexact}, the output of the first BMM is directly used by the softmax, and second BMM, without writing and reading from memory. Fusion significantly reduces memory overhead between ops. LIFE updates the statistics database accordingly when fusion is enabled, capturing the reduced memory usage and  dispatch calls.

% In LIFE, analytical model updates the statistics database accordingly when fusion is enabled. Fusion also reduces number of operator dispatch calls which are also monitored by LIFE. 

\subsubsection{Dynamic Shape Padding}
Padding arbitrary tensors to nearest supported operator shape is widely used to support dynamic shapes at runtime. In the decode phase, the KV cache incrementally increases by one token at a time. LIFE provides an analytical model of the attention operators to study padding during LLM decode. 

\subsection{Model Optimizations in LIFE}

\subsubsection{Model Quantization}
Quantization techniques enable parameter compression, resulting in low precision arithmetic and reduction in model parameter size. AWQ \cite{lin2024awqactivationawareweightquantization}, GPTQ \cite{frantar2022gptq} and QuaRot \cite{ashkboos2024quarotoutlierfree4bitinference} have proven that LLMs can be quantized with minimal loss of accuracy. Quantization requires dequantizing weights to higher precision within the compute operator before applying the linear affine transformation on the input activations. This behavior is modeled in LIFE's analytical model to reflect associated compute and memory overhead.

% in the the compute operator prior to performing the linear affine transformation on the input activations, which is models in LIFE's analytical models. 

\subsubsection{Attention mechanisms}
Different attention mechanisms, MHA, GQA, MQA or MLA are analytically modeled in LIFE. There is an emerging interest to convert existing LLMs trained with MHA/GQA/MQA to use MLA \cite{ji2025economicalinferenceenablingdeepseeks} as it reduces KV size and improves performance for long prompts. The analytical models in LIFE capture these attention mechanisms, including MLA, and a specific attention mechanism can be chosen using the config file. 

% which is also modeled in LIFE's analytical models. The attention mechanism is chosen using config file.   

\subsubsection{KV Cache compression}
KV compression  ~\cite{li2025surveylargelanguagemodel} reduces memory overhead of LLM inference \cite{hooper2025kvquant10millioncontext}. We consider two techniques for KV compression in the analytical model, (1) compression provided by the MLA operator (2) KV quantization to 4-bit or 8-bit. The resulting memory reduction is modeled by the LIFE  framework. Since KV quant requires dequantization during the attention operation, this overhead is also modeled in LIFE's analytical models. 

\subsubsection{LLM Chunked Prefill}

Chunked prefill splits large prompts into smaller chunks of equal sizes, computing the prefill for each chunk while reusing the KV cache from previous chunks \cite{agrawal2023sarathiefficientllminference}. Chunked prefill enables long prompts without the need for special hardware. LIFE models chunked prefill in its simulation framework. 
% LIFE's framework models it in it's simulation framework. The simulation script orchestrates simulation of chunked-prefill. 
The code snippet is shown in Fig.~\ref{life}.

\subsubsection{LoRA adaptation} 
LoRA adapters are merged with base model weights prior to computing the linear affine transformation in  Linear layers as shown in Equation-\ref{eq-5}. LIFE models the impact of LoRA on the Linear operator to characterize LoRA adaptation performed during inference for every token generation or at a one-time merge.  

\subsection{Simulation Scripts \& Statistical Database}
LIFE's simulation software runs the LLM analytical model to characterize inference workloads. The memory and compute utilization are accumulated in LIFE's statistics database. Because LIFE is dataset independent and does not use actual model weights or perform real inference, the time taken to characterize the workload is in the range of few seconds to minutes on a laptop, allowing for fast simulations. The configuration files configure the LLM analytical model to operate in required settings.

\section{Experimental Setup}
Below we enumerate the experiments performed with LIFE to characterize LLM inference workload and forecast the performance in a hardware and dataset agnostic way. 

\subsection{Types of Workload Experiments}
We organize LIFE's workload characterization abilities into two types: (1) operator-workload and (2) LLM workload. 
Furthermore, we consider both the workloads in both the prefill and decode stages of inference. 

\subsubsection{Operator workload}
Operator workload focuses on understanding operator workload in isolation, across prefill, decode and prefill-chunking. We study Linear layer with/without LoRA; BMM operation; different attentions in both prefill and decode modes. 
%The simulation script orchestrates the invocation of individual operators' analytical models to gather statistics metrics and forecast performance metrics. 

\subsubsection{LLM inference workload} 
LIFE also understands workload at the model level, invoking sequences of operators that constitute the model architecture. LIFE utilizes the model and software optimizations enumerated in its config file to generate different LLM scenarios for which LLM workload is studied. An example configuration file for Llama2-7B with MLA is shown in Appendix ~\ref{lst:llam-config-llama2}. 

\subsection{Metrics}
We categorize LIFE's metrics into two groups, (1) workload metrics and (2) performance metrics. 

\subsubsection{Workload Metrics} LIFE's simulation computes the following hardware agnostic workload metrics in the statistics database: \textbf{\textit{(1) Compute Operations}}, \textbf{\textit{(2) Total Memory read/write}} in bytes, \textbf{\textit{(3) KV Cache read/write}} in bytes, \textbf{\textit{(4) Number of Dispatch Calls}}. 
% \begin{enumerate}
%     \item \textbf{Compute Operations}
%     \item \textbf{Total Memory read/write} in bytes
%     \item \textbf{KV Cache read/write} in bytes
%     \item \textbf{Number of Dispatch Calls}    
% \end{enumerate}

\subsubsection {Performance Metrics} Analysis scripts in LIFE analyze the workload metrics with reference to a desired hardware system with TOPS, bandwidth (BW) (GBps) and operating efficiencies to forecast inference with the following performance metrics. 

To accurately forecast performance metrics, LIFE expects compute and memory efficiency of operator for specific shapes and extrapolates to other shapes. Efficiency on a hardware can be measured using unit tests of operators without running LLM inference workload. Alternatively, LIFE's analysis can forecast performance metrics for all ranges of efficiencies to give insights into what is possible on a given hardware configuration. 

\paragraph{\textbf{ (1) Time-To-First-Token (TTFT)}}
For a given hardware platform with known TOPS (TOPs/sec) with an average operator efficiency, the latency of the prefill phase is estimated as shown in ~Eq.\ref{eq-1} to ~Eq.\ref{eq-2}. 

    \begin{equation}\label{eq-1}
    t_{c} = \sum_{op} \left( \frac{TOPs_{op}}{\left({ec_{op}*TOPS} \right)} \right) + \sum_{op} t\_dispatch_{op}
    \end{equation}
    
    \begin{equation}\label{eq-1a}
    t_{m} = \sum_{op} \left( \frac{Mem_{op}}{\left({em_{op}*BW} \right)} \right) + \sum_{op} t\_dispatch_{op}
    \end{equation}

    \begin{equation}\label{eq-2}
    TTFT = \max \left( t_{c} ,\space t_{m} \right)
    \end{equation}

Specifically, \begin{math}t_{c}\end{math} is time taken to compute the operation and \begin{math} t_{m} \end{math} is time taken for memory access. $t\_dispatch_{op}$ is the dispatch latency of the operator. \begin{math} ec_{op}\end{math} and \begin{math} em_{op}\end{math} are the efficiencies of the compute operation and memory utilization of the operator $op$. When operator efficiency is 100\%, \begin{math} ec_{op}\end{math} and \begin{math} em_{op}\end{math} are both 1. 

% LIFE allows efficiency to be provided for specific operator shapes and extrapolates in between, if needed, for effective efficiency estimation.
% \begin{math} em_{avg}\end{math} is average memory utilization efficiency during the lifetime of LLM inference simulation.

\paragraph{\textbf{(2) Token-Per-Output-Token (TPOT})}
We observe that the $ t_{c} << t_{m} $ during the decode phase for the models and the conditions studied. Thus, we define TPOT to be solely dependent on $t_{m}$. Note, \begin{math} em_{avg}\end{math} is the average memory utilization efficiency during the lifetime of LLM inference simulation.
\begin{equation}\label{eq-3}
    TPOT = \sum_{ops} \frac{\left( BW * em_{op} \right)}{MEM_{op}} + \sum_{op} t\_dispatch_{op} 
\end{equation}    
\begin{equation}\label{eq-3a}
    TPOT = \frac{\left( BW * em_{avg} \right)}{MEM} + t\_dispatch_{total} 
\end{equation}
    
\paragraph{\textbf{(3) Token-Per-Second(TPS)}}
\begin{equation}\label{eq-4}
TPS = \left(1 / TPOT\right) tokens/sec
\end{equation}

\paragraph{\textbf{(4) LoRA adapter update time}}
$t_{lora}$ is the time taken to merge the product of LoRA adapters $A$, $B$ with the base weights of the corresponding linear layer.
\begin{equation}\label{eq-5}
t_{lora} = \sum_{linear} \left(W_{linear} + B_{linear}A_{linear} \right)
\end{equation}

\subsection{Models Studied}
We perform workload characterization of variants of Llama2-7B as shown in Table~\ref{tab:model-variants}. Each model variant is a combination of software and model optimizations.

\begin{table}[ht]
\centering
\resizebox{\columnwidth}{!}{%
\fontsize{10}{10}\selectfont
\begin{tabular}{|l|c|c|c|c|}
\hline
\textbf{Variant} & \textbf{Weights,} & \textbf{KV,}  & \textbf{LoRA} & \textbf{Fusion} \\
 & \textbf{Activations} & \textbf{Attention} &  &  \\
\hline
bf16-bf16 & bf16, bf16 & bf16, MHA & No & No \\
bf16-int4 & bf16, int4 & bf16, MHA & No & No \\
bf16-int4-fused & bf16, int4 & bf16, MHA & No & Yes \\
bf16-int4-kv4 & bf16, int4 & int4, MHA & No & Yes \\
bf16-int4-mla & bf16, int4 &  bf16, MLA & No & Yes \\
bf16-int4-lora & bf16, int4 & bf16, MHA & Yes & Yes \\
QuaRot-w4a4kv4 & int8, int4 & int4, MHA & No & Yes \\
fp16-fp16 & fp16, fp16 & fp16, MHA & No & No \\
\hline
\end{tabular}
}
\caption{Llama2-7B model variants studied}
\label{tab:model-variants}
\end{table}

\subsection{Verification Setup}
\label{sec:verif-setup}
To verify LIFE's forecasted performance metrics, we use three hardware platforms. (1) AMD Ryzen 9 HX 370 CPU ~\cite{amdryzenai} with 326.4 $GFLOPS$ and 240 $GBps$ bandwidth for both prefill and decode, and (2) AMD Ryzen AI Max+ 395 ~\cite{amdryzenaimax} with an overall capacity of 126 $TOPS$, 50 $TOPS$ on NPU for prefill and 256 $GBps$ bandwidth for iGPU for decode and (3) NVIDIA V100 GPU, with 126 $TOPS$ and 900 $GBps$ bandwidth ~\cite{nvidiav100}. For (1) and (3) we use Pytorch 2.6 \cite{NEURIPS2019_9015} and 4.49.0 version of HuggingFace transformers \cite{hftrans}. For (2), we use the RyzenAI hybrid-llm software ~\cite{amdryzenaillm}. LIFE simulation and analysis is run on setup (1).

\section{Results}
We perform in-depth characterization of LLM inference workload to collect workload metrics for different operating conditions and forecast the performance metrics for different hardware configurations and efficiencies. 

\subsection{Analysis of LLM Prefill}
\label{sec:llm-prefill}
\subsubsection{Operator distribution}
We characterize the prefill phase of Llama2-7B across varying prompt lengths and show detailed results for prompt length of 2048. LIFE's simulation provides workload metrics for all operators. Table~\ref{prefill-tops-vs-prompt} shows the distribution of compute operations across operators at diff prompt lengths. We observe that for shorter prompts, the Linear operator dominates compute usage, followed by BMM and Softmax in MHA. As the prompt length increases, the BMM operators in MHA significantly dominate the compute utilization. Note that at shorter prompts, BMM efficiency has lesser impact on TTFT compared to longer prompts.

\begin{table}[ht]
\centering
\resizebox{\columnwidth}{!}{%
\fontsize{10}{10}\selectfont
\begin{tabular}{|r|r|r|r|r|r|}
\hline
\textbf{Prompt} & \textbf{GEMM} & \textbf{BMM} & \textbf{Softmax} & \textbf{TOPs} & \textbf{KV (GB)}\\
\hline
256 & 99.0 \% & 1.0 \% & 0.0 \% & 3.42 & 0.1\\
1024 & 96.0 \% & 3.9 \% & 0.0 \% & 14.09 & 0.5\\
2048 & 92.4 \% & 7.5 \% & 0.1 \% & 29.29 & 1.0\\
4096 & 85.9 \% & 14.0 \% & 0.2 \% & 63.04 & 2.0\\
8192  & 75.2 \% & 24.5 \% & 0.3 \% & 143.87 & 4.0\\
16384  & 60.3 \% & 39.1 \% & 0.5 \% & 358.94 & 8.0\\
32768  & 43.2 \% & 56.0 \% & 0.7 \% & 1002.67 & 16.0\\
65536  & 27.5 \% & 71.6 \% & 0.8 \% & 3144.41 & 32.0 \\
\hline
\end{tabular}
}
\caption{TOPs vs Prompt length for Llama2-7B bf16-bf16 model in Prefill}
\label{prefill-tops-vs-prompt}
\end{table}

\begin{figure*}[htbp]
\centerline{\includegraphics[width=\textwidth]{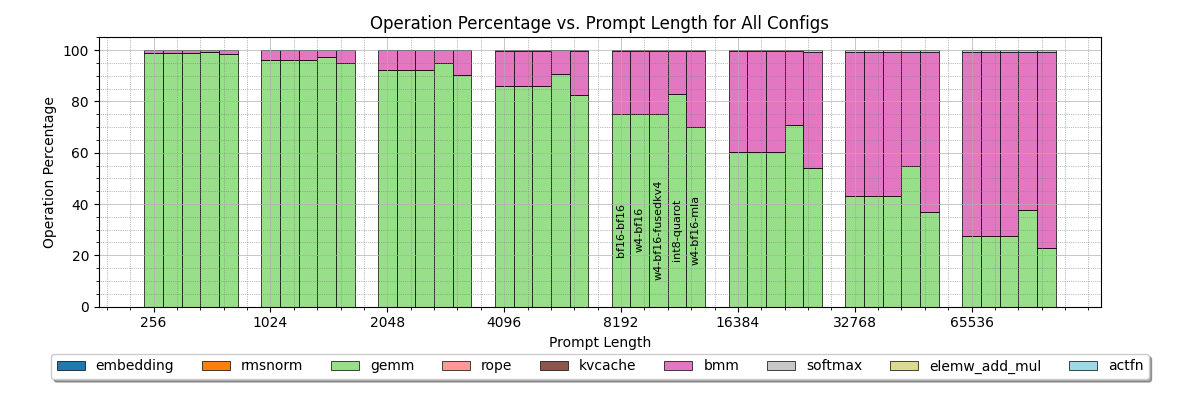}}
\caption{Prompt Length vs TOPs for Llama2-7B variants in Prefill: Each bar represents a breakdown of TOPs across different operators in that variant, (1) bf16-bf16 (2) bf16-int4 (3) bf16-int4-kv4 (4) QuaRot int8-int4 with Hadamard (5) bf16-int4-mla with Q, KV rank of 128. MLA with low rank adapters multiplied online at runtime }
\label{fig-prefill-models}
\end{figure*}

We further analyze compute workload distribution across 
the model variants described in Table~\ref{tab:model-variants}. Fig.~\ref{fig-prefill-models} shows compute TOPs breakdown by operator for each model variant. We observe that despite applying model optimizations, compute complexity remains the primary bottleneck during the prefill phase. %Moreover, when MLA is enabled, prefill latency depends on ranks of MLA and whether the adapters are merged with Q, KV matrices ahead of time or at runtime. 

% The study reveals that irrespective of applying model optimization strategies, the compute complexity is primary bottleneck for prefill phase. Moreover, when MLA is enabled, prefill latency depends on ranks of MLA and whether the adapters are merged with Q, KV matrices ahead of time or at runtime. 

We also examine how compute and memory usage vary with (1) software optimizations (2) model quantization and (3) model optimizations with KV cache. Table~\ref{table:llama2-7b} summarizes these results. We see that the TOPs remain largely unchanged across quantization and optimization techniques, but that the memory utilization varies significantly.
 
 \begin{table}[ht]
\centering
\resizebox{\columnwidth}{!}{%
\fontsize{10}{10}\selectfont
\begin{tabular}{|l|c|c|c|c|c|c|}
\hline
\textbf{Model} & \textbf{Prompt} & \textbf{TOPs} & \textbf{MemRD} & \textbf{MemWR} & \textbf{KV} \\
 &  \textbf{Length} &   & \textbf{(GB)} & \textbf{(GB)} & \textbf{(GB)} \\
\hline
\textbf{bf16-bf16} & 2048 & 29.2941 & 43.5 & 29.0 & 1 \\
\textbf{bf16-int4} & 2048 & 29.3074 & 34.4 & 29.0 & 1 \\
\textbf{bf16-int4-kv4} & 2048 & 29.3079 & 10.1 & 4.4 & 0.25 \\
\textbf{bf16-bf16} & 4096 & 63.0379 & 106.4 & 90.1 & 2 \\
\textbf{bf16-int4} & 4096 & 63.0511 & 97.3 & 90.1 & 2 \\
\textbf{bf16-int4-kv4} & 4096 & 63.0522 & 16.8 & 8.8 & 0.5 \\
\hline
\end{tabular}
}
\caption{Model workload metrics for Llama2-7B variants}
\label{table:llama2-7b}
\end{table}

\subsubsection{Forecasting Prefill Performance}
We use LIFE's analysis scripts to forecast performance metrics using the workload metrics for two model variants, bf16-int4 and bf16-int4-kv4, focusing on a 4096 prompt length. The simulation evaluates the impact of compute and memory efficiency across different hardware configurations ranging from 
from 10-100 TOPS of compute and from 10-100GBps of peak bandwidth. For the 100 hardware configurations, we measure \begin{math}t_{c}\end{math} and \begin{math}t_{m}\end{math} using Equation~\ref{eq-1}-~\ref{eq-2}, which loosely correlates with arithmetic intensity. 

\begin{figure*}[htbp]
\centerline{\includegraphics[width=\textwidth]{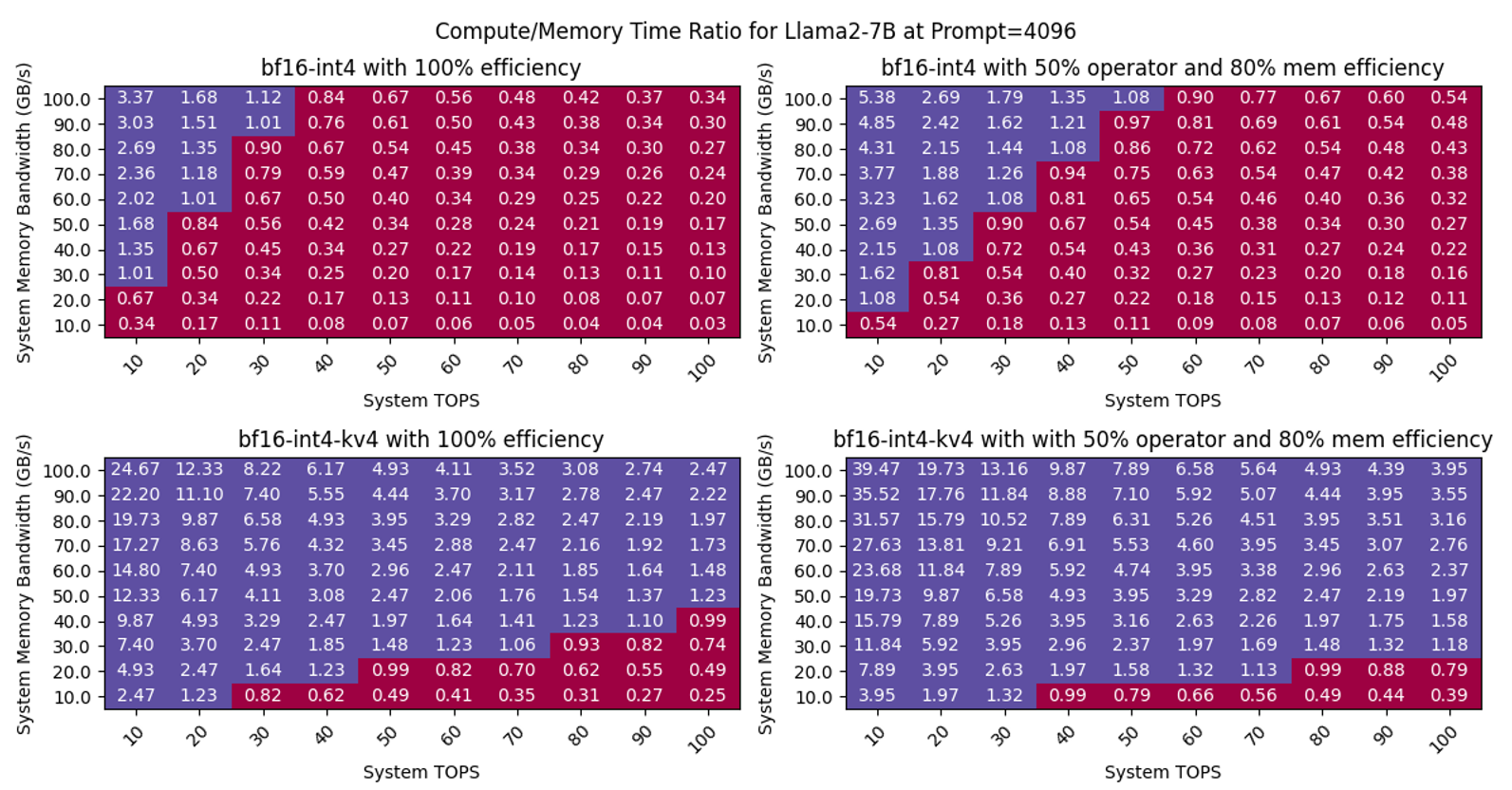}}
\caption{Llama2-7B: $t_{c}/t_{m}$ for different hardware configurations, compute and memory efficiency}
\label{fig-prefill-llama2-bf16-grid}
\end{figure*}

~Fig.~\ref{fig-prefill-llama2-bf16-grid}, shows a grid of these ratios. When \begin{math}\left(t_{c}/t_{m}\right) > 1\end{math}, the TTFT is limited by compute bound and when \begin{math}\left(t_{c}/t_{m}\right) < 1\end{math}, TTFT is bandwidth bound. For the bf16-int4 model at 100\% efficiency, prefill is predominately memory-bound due to high memory read/writes (~Fig.~\ref{fig-prefill-llama2-bf16-grid} top-left). Reducing compute to 50\% and memory efficiency to 80\% shifts the performance profile, altering the \begin{math}\left(t_{c}/t_{m}\right)\end{math} balance (~Fig.~\ref{fig-prefill-llama2-bf16-grid} top-right. In this case, although the total TOPS decreased, the drop in memory efficiency had a greater impact, shifting the \begin{math}\left(t_{c}/t_{m}\right) \end{math} ratio, and altering the performance bottleneck. We repeat the same analyses with the bf16-int4-kv4 model variant (bottom left and right) and observe a significant difference in \begin{math}\left(t_{c}/t_{m}\right) \end{math}, highlighting the impact of KV compression on memory. Note that if \begin{math}\left(t_{c}/t_{m}\right) >1 \end{math} model optimization techniques do not help with TTFT.
Overall, the results from Fig.~\ref{fig-prefill-llama2-bf16-grid} reveal that traditional roofline analyses are insufficient for forecasting TTFT in LLMs, as operator efficiencies vary greatly during inference. Single roofline models do not capture all dynamisms adequately; hardware efficiency varies with LLM operating conditions.

\begin{figure}[htbp]
\centerline{\includegraphics[width=\columnwidth]{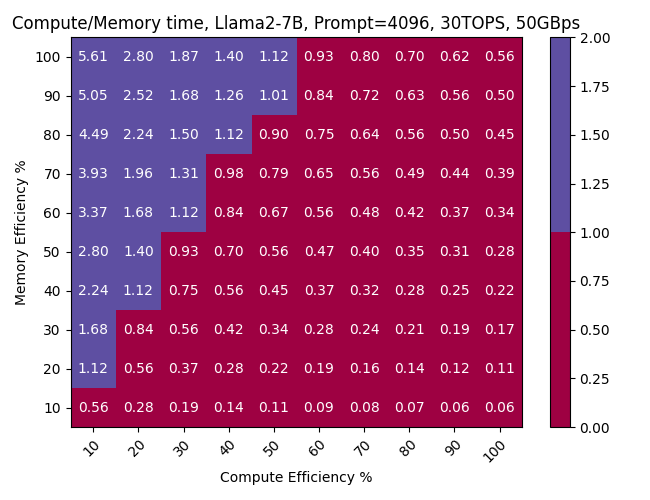}}
\caption{Llama2-7B tc/tm on 50TOPs, 30GBps vs compute and bandwidth efficiencies}
\label{fig-30tops-50gbps}
\end{figure}

We further analyze a single configuration within the 100 hardware configurations--30 TOPS and 50 GBps--to study how varying compute and memory efficiency affect prefill latency. As shown in Fig.~\ref{fig-30tops-50gbps}, TTFT can be accurately predicted when the compute and hardware efficiencies and how they vary in time and operator space is understood. 

% We further analyze a single datapoint out of the 100, for a system with 30 TOPS and 50 GBps to study effect of compute and memory efficiency on the prefill latency. Fig.~\ref{fig-30tops-50gbps} shows the nuanced effects compute and memory efficiencies have on the performance. We observe that for a given system, we can accurately forecast TTFT if we understand the compute and efficiencies and how they vary in time and operator space. 

\subsubsection{Prefill Forecast verification}

We verified LIFE's forecasting and characterization on two hardware setups. ~Table~\ref{tab:llama-2-7b-decode} compares measured and forecasted performance. On setup 1 (CPU), we measure the bf16-bf16 variant. Interestingly, as the matrix dimensions increase, compute efficiency drops--contrary to expectations--due to increased dispatch calls and cache pressure which raise both $t_{m}$ and $t_{c}$. On setup 2 (NPU), we measure the bf16-int4 variant. On setup (3) we measure fp16-fp16 variant. The absolute TTFT is much shorter compared to setup 1 due to faster hardware and quantized weights. We also observe that compute efficiency improves with longer prompts, indicating better NPU efficiency on longer prompts. In all three cases, LIFE's hardware and dataset agnostic forecasts closely match measurements on real hardware. All three hardwares are vastly different CPU, NPU, iGPU and data center GPU. 

% It is interesting to observe that as the matrix dimensions increase, the efficiency of computation drops, which is counterintuitive. $t_{m}$ increases along with $t_{c}$, number of dispatch calls increase as the large matrices cannot fit into the available cache of CPU, thereby reducing the overall efficiency. On setup 2 with NPU, we measure performance of bf16-int4 variant. The absolute TTFT is much smaller compared to setup 1 due to faster NPU, and quantized weights. We also observe efficiency of compute increases as prompt length increases. This shows that the efficiency of NPU is much higher for larger prompts. In both cases, we verify that LIFE's hardware and dataset agnostic forecasted performance is close to measurements on real hardware.

\begin{table}[htbp]
\caption{Forecast vs Measurements - Prefill}
\centering
\resizebox{\columnwidth}{!}{
\fontsize{10}{10}\selectfont
\begin{tabular}{|l|c|c|c|c|c|}
\hline
\textbf{Prompt} & \textbf{TOPs} &  
\multicolumn{2}{c|}{\textbf{Forecast TTFT}} & 
\textbf{Measured} & \textbf{Measured} \\
\textbf{Length}& & \textbf{100\%} & \textbf{50\%} & \textbf{TTFT} & \textbf{Efficiency} \\
& & \textbf{efficiency} & \textbf{efficiency} & & \\
\hline
\multicolumn{6}{|c|}{\textbf{AMD Ryzen 9 HX 370 CPU: bf16-bf16}} \\
\hline
32 & 0.42 &  1.30 & 2.60 & 1.85 & 70.3 \% \\
64 & 0.85 &  2.61 & 5.21 & 3.34 & 77.9 \% \\
128 & 1.70 & 5.21 & 10.42 & 6.72 & 77.5 \% \\
256 & 3.42 & 10.48 & 20.96 & 14.61 & 71.7 \% \\
512 & 6.91 & 21.17 & 42.34 & 31.03 &  68.2 \% \\
1024 & 14.09 & 43.17 & 84.34 & 72.99 & 59.1 \% \\
2048 & 29.29 & 89.74 & 179.47 & 186.15 & 48.2 \% \\
\hline
\multicolumn{6}{|c|}{\textbf{AMD Ryzen AI Max+ 395 NPU: bf16-int4}} \\
\hline
128 & 1.70 & 0.04 & 0.07 & 0.3 &  11.3\% \\
1536 & 21.55 & 0.43 & 0.86 & 1.8 &  23.9\% \\
\hline
\multicolumn{6}{|c|}{\textbf{NVIDIA V100 GPU: fp16-fp16}} \\
\hline
512 & 6.91 & 0.06 & 0.11 & 0.11 & 50.3 \% \\
1024 & 14.09 & 0.11 & 0.22 & 0.2 & 56.3 \% \\
2048 & 29.29 & 0.23 & 0.47 & 0.4 & 58.6 \% \\
\hline
\end{tabular} 
}
\label{tab:llama-2-7b-measured}
\end{table}

\subsection{LLM Prefill with chunking}
\begin{figure*}[htbp]
\centerline{\includegraphics[width=\textwidth]{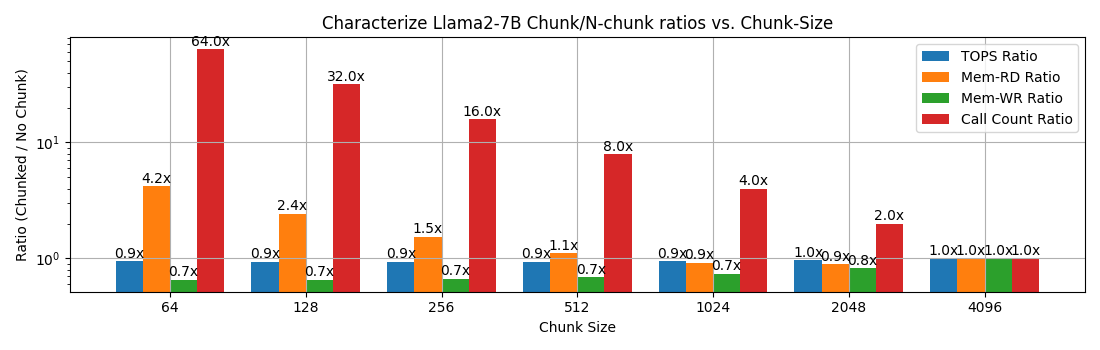}}
\caption{Llama2-7B: Ratio of metrics chunked-prefill/no chunked-prefill at prompt=4096}
\label{fig-prefill-chunk}
\end{figure*}

Chunked-prefill is an effective technique for supporting long prompts that exceed system limits. We study prefill-chunking for the Llama2-7B bf16-bf16 variant for a prompt length of 4096 tokens with different chunk sizes.  Fig.~\ref{fig-prefill-chunk} shows workload metric ratios relative to no chunking. We see that smaller chunk sizes increase memory pressure (orange bar), but as long as $t_{c}>t_{m}$, the process remains compute bound and chunked prefill does not slow down compared to regular prefill. 
We find that the efficiency of memory utilization must be kept high to keep the prefill compute bound. This means, to forecast accurate TTFT for prefill chunking, analyses shown in Sec. \ref{sec:llm-prefill} must be done. Additionally, the number of dispatch calls also increases by 64x for smallest chunk-size (red bar). However, given the operator dispatch latency is several orders of magnitude smaller than the compute and memory latency, it has no impact to LLM prefill performance. We observe that the prefill chunk size inversely affects memory utilized, but compute load change minimally.

 % We study prefill-chunking for Llama2-7B bf16-bf16 variant for a prompt length of 4096 tokens with different chunk sizes. LIFE's simulation setup gives us workload metrics. We plot the workload metrics ratio with respect to no chunking in Fig.~\ref{fig-prefill-chunk}. We observe that the smaller the chunk size, higher the impact to memory (orange). Thus, when chunked-prefill is implemented, as long as $t_{c}>t_{m}$, it is still compute bound and chunked prefill does not slow down compared to regular prefill. We find that the efficiency of memory utilization must be kept high to keep the prefill compute bound. This means, to forecast accurate TTFT, thorough analysis shown in earlier section (Fig.~\ref{fig-prefill-llama2-bf16-grid} and Fig.~\ref{fig-30tops-50gbps}) must be done, as the plots change.   

\subsection{Analysis of LLM Decode}
The ~Table~\ref{tab:llama-2-7b-decode} shows the workload metrics, GOPs and total memory utilized by the Llama2-7B variants during decode. We see that compute workload is three orders of magnitude lower than in prefill phase in Table~\ref{prefill-tops-vs-prompt} . Although the GOPs increase by ~2x in the int4 model due to dequantization ops, the total compute is few GOPs. However, memory usage is substantial and grows with token generation. We show that the biggest factor for TPS improvement during decode is driven by int4 quantization. Additionally, while KV compression has minimal effects on small prompts, its impact is prominent at higher prompts.

\begin{table}[htbp]
\caption{Analysis of Llama2-7B variants' GOPs, Memory in decode phase}
\centering
\resizebox{\columnwidth}{!}{
\fontsize{10}{10}\selectfont
\begin{tabular}{|l|c|c|c|c|c|c|}
\hline
\textbf{Prompt} & 
\multicolumn{3}{c|}{\textbf{GOPs}} & 
\multicolumn{3}{c|}{\textbf{Memory (GB)}} \\
& \textbf{bf16} & \textbf{bf16} & \textbf{bf16-} & \textbf{bf16} & \textbf{bf16} & \textbf{bf16-} \\
&  \textbf{-bf16} & \textbf{-int4} & \textbf{int4-kv4}  & \textbf{-bf16} & \textbf{-int4} & \textbf{int4-kv4} \\
\hline
32 & 13.34 & 26.55 & 26.61 & 12.85 & 3.74 & 3.55 \\
64 & 13.36 & 26.57 & 26.64 & 12.88	& 3.77 & 3.57 \\
128 & 13.39	& 25.60 & 26.69	& 12.94 & 3.83 & 3.59 \\
256 & 13.46	& 26.67 & 26.79 & 13.07 & 3.96 & 3.59 \\
512 & 13.59	& 26.81 & 26.99 & 13.32 & 4.21 & 3.64 \\
1024 & 13.86 & 27.08 & 27.40 & 13.82 & 4.71 & 3.73 \\
2048 & 14.41 & 27.62 & 28.21 & 14.83 &	5.72 &	3.92 \\
\hline 
\end{tabular} 
}
\label{tab:llama-2-7b-decode}
\end{table}
% The compute workload for the three variants is found to be 3 orders of magnitude less than that in prefill stage. Even though the GOPs increase by almost 2x in int4 variants due to dequantize step the absolute compute required is still in few GOPS, much less than prefill. However, the memory utilized by the models is significantly high and increases as more new tokens are generated. We show that the biggest factor for TPS improvement during decode comes from int4 quantization. For smaller prompts, KV compression does not add much benefit but the impact is prominent at higher prompts.  

\subsubsection{Operator dispatch latency}
There is a relatively low latency cost for every operator dispatch call. During the decode stage, the dispatch call latency adds to the TPOT as the workload and impacts TPOT. LIFE's simulation framework gathers dispatch calls as a workload metric. ~Table.\ref{tab:decode-dispach-calls} shows number of dispatch calls for different models. Operator fusion reduces number of dispatch calls. 

\begin{table}[htbp]
\caption{Analysis of Dispatch calls during decode}
\centering
%\resizebox{0.7\columnwidth}{!}{
\fontsize{10}{10}\selectfont
\begin{tabular}{|l|c|}
\hline
\textbf{Model} & \textbf{Dispatch Calls} \\  
\hline 
Llama2-7B-int4 & 611 \\
DeepSeek-Qwen2.5-1.5B-int4 & 535 \\
DeepSeekv2-Lite-int4 & 516 \\
Gemma2-2B-int4 & 497 \\
Phi4-3.8B-int4 & 515 \\
Qwen3-14B-int4 & 763 \\
Qwen3-32B-int4 & 1219 \\
\hline
\end{tabular} 
\label{tab:decode-dispach-calls}
\end{table}

\subsubsection {Long context and generation}
\begin{figure*}[htbp]
\centerline{\includegraphics[width=\textwidth]{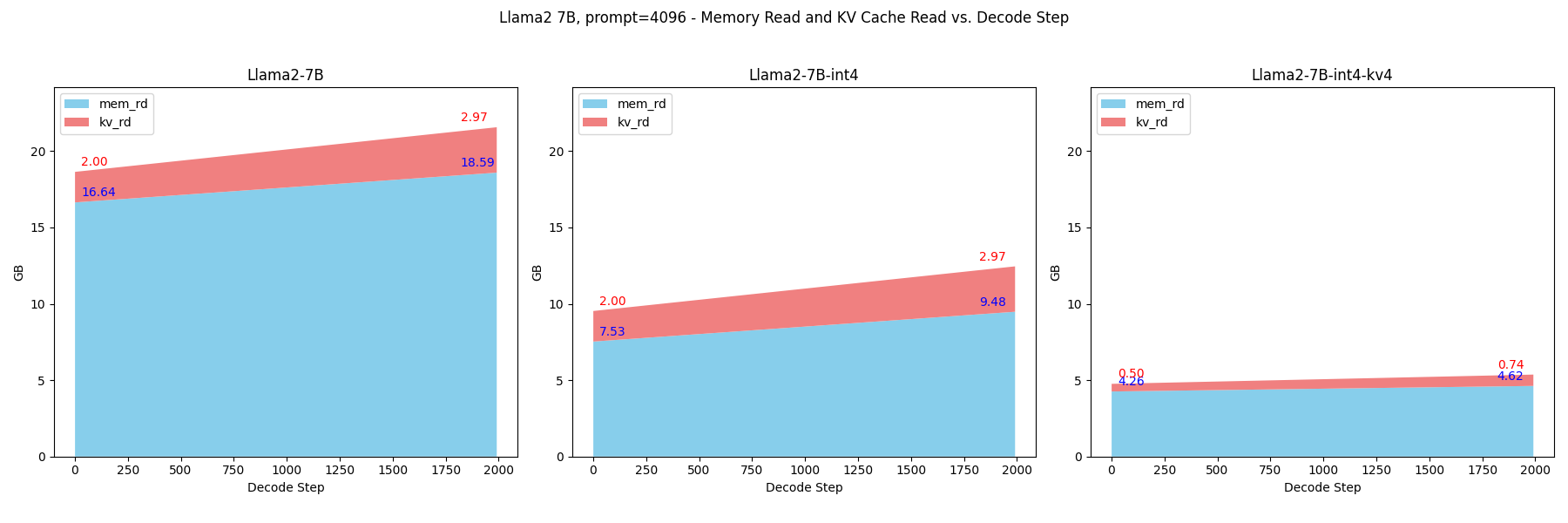}}
\caption{Llama2-7B Memory RD during Decode for prompt=4096 and 2000 new tokens: (left) bf16-bf16 (middle) bf16-int4 (right) bf16-int4-kv4}
\label{fig-decode-long}
\end{figure*}

As text generation progresses, KV cache accumulates and the overall memory utilized to generate a single token increases. Quantifying this metric is critical to understand how the TPOT and TPS change over time for a given prompt length. Fig.~\ref{fig-decode-long} shows the memory utilized during model decode starting from a 4K initial prompt length. The X axis shows new tokens generated while the Y axis shows memory consumed. From Equations.\ref{eq-1}-\ref{eq-4}, TPOT is directly proportional to memory read. We observe that without KV compression, TPS drops as high as 50\% and 26\% for smaller and longer prompts respectively with same bandwidth efficiency. With KV compression, TPS does not drop more than 10\% for any prompt length.

\begin{table}[htbp]
\caption{Analysis of Llama2-7B Memory during Decode with 4K prompt and 2000 new tokens}
\centering
\resizebox{\columnwidth}{!}{
\fontsize{10}{10}\selectfont
\begin{tabular}{|l|l|r|r|r|}
\hline
\textbf{Prompt} & \textbf{Llama2-7B} & \textbf{Mem(GB)} & \textbf{Mem(GB)} & \textbf{Mem last/} \\
\textbf{Length} & \textbf{variant} & \textbf{1st token} &\textbf{last token} & \textbf{Mem 1st}\\
\hline
128 & bf16-bf16 & 12.75 & 14.71 & 1.15x\\
128 & bf16-int4 & 3.65 & 5.60 & 1.53x \\
128 & bf16-int4-kv4 & 3.53 & 3.90 & 1.10x \\
\hline
4096 & bf16-bf16 & 16.66 & 18.62 & 1.18x \\
4096 & bf16-int4 & 7.55 & 9.51 & 1.26x \\
4096 & bf16-int4-kv4 & 4.26 & 4.60 & 1.08x \\
\hline 
\end{tabular} 
}
\label{table:llama2-7b-long-context}
\end{table}

\subsubsection{Decode Forecast Verification}
LIFE's analysis software computes \begin{math} t_{m} \end{math} and $t\_dispatch_{op}$ and performance metrics with Equation.\ref{eq-1} through ~Equation.\ref{eq-4}. We first forecast the TPS for a given efficiency and then measure the TPS on the actual hardware on both setups 1 and 2 describe in Section.~ref{sec:verif-setup}. The results are shown in ~Table.~\ref{tab:llama-2-7b-decode-measured}. We observe that the forecasted TPS is comparable to measured TPS on two independent hardware platforms. This verified LIFE's hardware agnostic and dataset independent performance forecasting.

\begin{table}[htbp]
\caption{Forecast vs Measurements Llama2-7B Decode}
\centering
\resizebox{\columnwidth}{!}{
\fontsize{10}{10}\selectfont
\begin{tabular}{|l|c|c|c|c|c|}
\hline
\textbf{Prompt} & 
\multicolumn{3}{c|}{\textbf{Measured}} & 
\multicolumn{2}{c|}{\textbf{Forecast}} \\
\textbf{length} & \textbf{TPOT(ms)} & \textbf{TPS} & \textbf{Efficiency} & \textbf{Efficiency} & \textbf{TPS} \\
\hline 
\multicolumn{6}{|c|}{\textbf{AMD Ryzen 9 HX 370 CPU: bf16-bf16}} \\
\hline
32 & 629 & 1.59 & 9\% & 10\% & 1.87 \\
64 & 608 & 1.64 & 9\% & 10\% & 1.86 \\
128 & 769 & 1.30 & 7\& & 10\% & 1.85 \\
256 & 574 & 1.74 & 10\% & 10\% & 1.84 \\
512 & 903 & 1.11 & 6\% & 10\% & 1.80 \\
1024 & 1152 & 0.87 & 5\% & 10\% & 1.74 \\
2048 & 2203 & 0.45 & 3\% & 10\% & 1.62 \\
\hline 
\multicolumn{6}{|c|}{\textbf{AMD Ryzen AI Max+ 395 iGPU: bf16-int4}} \\
\hline
128 & 28.7 & 34.5 & 52.5\% & 50\% & 33.4 \\
1536 & 30.5 & 32.8 & 52.5\% & 50\% & 27.2 \\
\hline
\multicolumn{6}{|c|}{\textbf{NVIDIA V100 GPU: fp16-fp16}} \\
\hline
512 & 25 & 40.0 & 60\% & 50\% & 32.6 \\
1024 & 27 & 36.9 & 57\% & 50\% & 30.3 \\
2048 & 31 & 32.1 & 51\% & 50\% & 26.7 \\
\hline
\end{tabular} 
}
\label{tab:llama-2-7b-decode-measured}
\end{table}

\subsection{Analysis of Attention Mechanisms}
We use LIFE's simulation framework to characterize all attention mechanisms during decode phase and compare them. LIFE gathers the statistics for 2000 consecutive output tokens starting from a prompt length of 8192. Table.\ref{tab:attn-compare} shows the results of comparison. We observe that MLA consumes more memory during decode than MQA and GQA for long prompts with online computation of low rank adapters within Q and KV Linear layers of the MLA. However, when the KV cache is compressed along with MLA, memory consumed by MLA reduces to almost as much as GQA. In all cases, MQA consumes least memory. For long prompt, replacing MHA with MLA reduces memory consumption by almost 50\%. The investigation conclusively shows that GQA with KV cache compressed is comparable to MLA in memory utilization. 

\begin{table}[htbp]
\caption{Attention mechanism memory comparison during LLM decode}
\centering
\resizebox{\columnwidth}{!}{
\fontsize{10}{10}\selectfont
\begin{tabular}{|l|c|c|}
\hline
\textbf{Mode} & 
\textbf{Mem[MB]-1st token} & \textbf{Mem[MB]-2000th token} \\
& \textbf{MHA/GQA/MQA/MLA} & \textbf{MHA/GQA/MQA/MLA}  \\
\hline
Eager & 388 / 244 / 202 / 344 & 450 / 283 / 234 / 415 \\
Fused & 322 / 178 / 136 / 278 & 368 / 201 / 152 / 333 \\
Fused-KV8 & 226 / 130 / 102 / 166 & 249 / 141 / 110 / 193 \\
Fused-KV4 & 178 / 106 / 85 / 110 & 189 / 111 / 89 / 124 \\
\hline 
\end{tabular} 
}
\label{tab:attn-compare}
\end{table}

\subsubsection{Analysis of MHA Efficiency}
\begin{figure}[htbp]
\centerline{\includegraphics[width=\columnwidth]{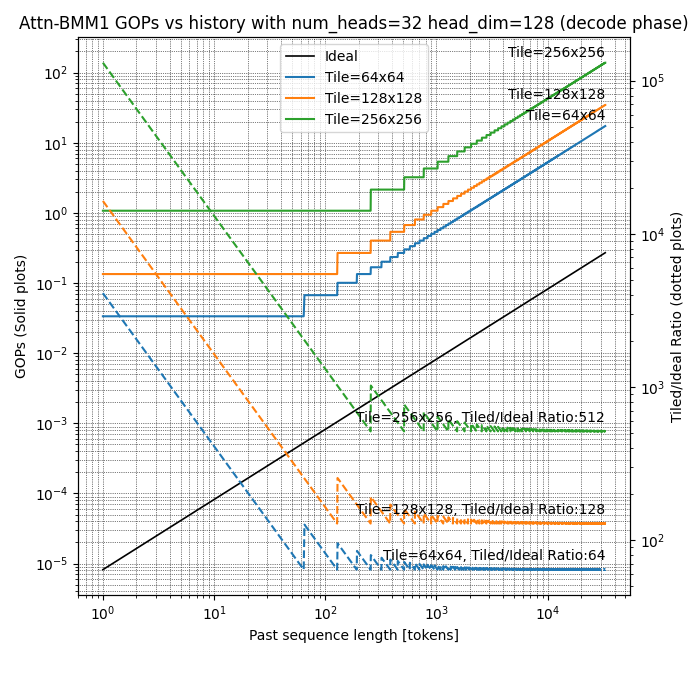}}
\caption{BMM Tiling efficiency}
\label{fig-decode-bmm-mha-efficiency}
\end{figure}

The decode phase increases the KV cache by one token for every new token generated, consequently incrementing the inputs to the BMM operator in MHA by one. BMM is often performed by padding the inputs to closest supported tiled implementation in hardware. The compute efficiency of BMM kernel drops due to under utilization. As the inputs increase, the padding required reduces, eventually BMM operates at maximum efficiency, and as the inputs increase further, the utilization for the last BMM tile drops and this repeats. With LIFE framework, we characterize and analyze the BMM operator in this specific operating condition to quantify the efficiency for various tile sizes. ~Fig.\ref{fig-decode-bmm-mha-efficiency} shows the results. The y-axis is compute ops. The solid black line is ideal compute. We observe that for a tile size of 64 (blue), the ideal BMM compute (dotted blue line) and the tiled BMM compute (solid blue line) differ by 1000x as sequence length increases (right). The different color lines show that irrespective of tile size selection, efficiency drops drastically as sequence length increases. 

The sawtooth plot of BMM efficiency variation in decode show that the average efficiency reaches an asymptote as the number of new tokens generated increases. The asymptote defines the average BMM efficiency for long token generation with large KV cache. Thus, for long prompts, efficiency of BMM operator becomes critical. Forecasting TPS with varying MHA operator efficiency is a challenge due varying efficiencies.

\subsection{Analysis of LoRA adaptation}
\begin{figure}[htbp]
\centerline{\includegraphics[width=\columnwidth]{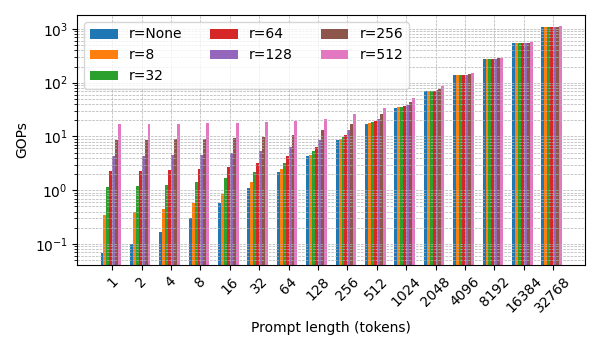}}
\caption{LoRA Linear: Ops vs prompt-size vs LoRA rank}
\label{fig-gemm-lora}
\end{figure}

We used the LIFE framework to investigate impact of LoRA adapter on TTFT. Fig.~\ref{fig-gemm-lora} shows compute ops for a single GEMM of size 4096x4096, appended with a LoRA adapter. We first analyze the impact on a single GEMM operator with LoRA adapter. We observe that for LoRA adapter merge inline with matmul, the compute operations increase vastly compared to the baseline of no LoRA adapters as the rank increases. This is shown in the first nine sets of bars on the plot. As the prompt length increases, compute required for adapter merge is much lesser than GEMM. As shown in the ~Fig.~\ref{fig-prefill-llama2-bf16-grid}, absolute TOPs for TTFT phase for smaller prompts is order(s) of magnitude lower than for 2K or more prompt. We conclude that if the LoRA adapter is inline in the matmul op for every single GEMM call, TTFT of smaller prompts increases by almost 2x compared to longer prompts.

Secondly we analyze the total TOPs required to do a complete model update with LoRA adapters on all 7 matmuls of Llama2-7B. A full model LoRA update requires considerable amount of compute as rank increases. Compare to compute for prefill in Table~\ref{prefill-tops-vs-prompt}, TTFT for a prompt length of 256 requires 3.42 TOPs. LoRA update is almost half of that, at 1.67 TOPs for r=128. Therefore, continuous LoRA update during inference impacts TTFT by more than 50\% for smaller prompts, where as the impact is lesser for larger prompts. If the update is done once, ahead of time, TTFT does not get affected due to LoRA.

\begin{table}[htbp]
\caption{Llama2-7B LoRA update compute overhead}
\centering
\resizebox{\columnwidth}{!}{
\fontsize{10}{10}\selectfont
\begin{tabular}{|l|c|c|c|c|c|c|}
\hline
\textbf{Layer} & \textbf{K} & \textbf{N} & \textbf{r=16} & \textbf{r=32} & \textbf{r=64} & \textbf{r=128} \\
\hline
q\_proj & 4096 & 4096 & 0.6 & 1.1 & 2.2 & 4.3 \\
k\_proj & 4096 & 4096 & 0.6 & 1.1 & 2.2 & 4.3 \\
v\_proj & 4096 & 4096 & 0.6 & 1.1 & 2.2 & 4.3 \\
o\_proj & 4096 & 4096 & 0.6 & 1.1 & 2.2 & 4.3 \\
gate\_proj & 4096 & 11008 & 1.5 & 3.0 & 5.9 & 11.6 \\
up\_proj & 4096 & 11008 &  1.5 & 3.0 & 5.9 & 11.6\\
down\_proj & 11008 & 4096 & 1.5 & 3.0 & 5.9 & 11.6\\
\hline
Total (x32)  &  - & - & 220.2 & 427.4 & 841.9 & 1670.8\\
\hline 
\end{tabular} 
}
\label{tab:llama-2-7b-decode-measured-lora}
\end{table}

\section{Related Work}
Forecasting LLM performance is of paramount importance due increasing LLM sizes and new hardware. ASTRA \cite{9238637} and other simulators like ~\cite{Lee_2025}, ~\cite{Cho_2024} have shown ways to simulate the workload on either specific networks, or specific hardware architectures such as GPU. Neusight ~\cite{Lee_2025} specifically targets GPU hardware performance ASTRA focusses on large scale training performance. Vidur ~\cite{agrawal2024vidurlargescalesimulationframework} presents a large scale simulation framework that predicts the best deployment configuration to make inference efficient but does not address the causes of performance bottlenecks and impact of efficiency. PyTorch profiler \cite{NEURIPS2019_9015} gives some insights into workload but it lacks efficiency based performance forecasting. \cite{kundu2024performancemodelingworkloadanalysis} is another interesting work that targets modeling LLM performance but only for GPU. We found that a fundamental hardware and dataset agnostic analytical model of LLM inference like \textbf{LIFE} and performance forecasting through the lens of compute or memory efficiency has not been thoroughly studied. 

\section{Conclusion}
Forecasting LLM performance using efficiency based analytical model is essential to quantitatively evaluate dynamically varying LLM inference workload. This paper delved into the understanding of the inherent dynamism of the LLM inference workload and performance forecasting through the lens of system efficiency. Our analysis using LIFE demonstrated that the compute and memory efficiency play a critical role in performance, beyond a simple roofline model. We showed the percentage impact to TTFT in prefill phase, TPS  during decode phase for various software and model optimizations, for varying operating conditions of short and long prompts. We also analyzed and highlighted the role of BMM efficiency for long prompts and GEMM for LoRA adaptation. We showcased critical bottlenecks of the workload and how the analysis can be leveraged to forecast performance variation on hardware. We further showed how widely used LoRA adapters have impact on TTFT at different prompt lengths. Finally, we compared LIFE's forecasted performance metrics to real measurements on AMD CPU, NPU, iGPU and NVIDIA GPU and showed accuracy of our forecasting. We show that LIFE's framework can be applied to map LLMs on to any hardware. While we showcase our study on dense LLMs, extending this to Vision Language Models (VLMs), Mixture-of-Experts (MoEs) and Speculative Decoding is left for future exploration.

\bibliographystyle{IEEEtranS}
\bibliography{reference}

\section{Appendix}
\subsection{Analytical Model of Linear} \label{appendix-a}

\begin{lstlisting}[language=Python,label={lst:gemm-model}]
def gemm(cls, opconfig:Dict={}, adapter_only:bool=False, strict:bool=False) -> Tuple:
    (m, k) = opconfig["shape_a"]
    (_, n) = opconfig["shape_b"]
    g = opconfig["grpsize"]
    opcount = m * k * n * 2 - (m * n)
    mem_rd  = (m * k) * cls.calc_nbytes(opconfig["dtype_a"])
    mem_wr =  (m * n) * cls.calc_nbytes(opconfig["dtype_out"])
    mem_rd_params = (k * n) * cls.calc_nbytes(opconfig["dtype_b"])
    if opconfig["bias"]:
        opcount += m * n 
        mem_rd_params += n * cls.calc_nbytes(opconfig["dtype_a"])
    if (opconfig["dtype_b"] == "int4"): # per group
        # dequant 
        opcount += (k * n) * 2
        mem_rd_params += ((k//g)*n) * cls.calc_nbytes(opconfig["dtype_a"]) # scale
        mem_rd_params += ((k//g)*n) * cls.calc_nbytes(opconfig["dtype_b"]) # zero 
    # if LoRA 
    if ("lora_rank" in opconfig.keys()):
        if (opconfig["lora_rank"] is not None):
            mem_rd_params += (k * opconfig["lora_rank"]) * cls.calc_nbytes(opconfig["dtype_lora"])
            mem_rd_params += (opconfig["lora_rank"] * n) * cls.calc_nbytes(opconfig["dtype_lora"])
            opcount += (k * opconfig["lora_rank"] * n) * 2 # lora A@B
            opcount += (k * n) # addition of adapter to original wt matrix
    cls.update_stats_ops("gemm", opcount, mem_rd, mem_wr, opconfig["mode"])
    cls.update_stats_ops("gemm", 0, mem_rd_params, 0, "eager")
    return  (m, n)
\end{lstlisting}

\subsection{LLM configuration file with MLA}
\begin{lstlisting}[language=Python,label={lst:llam-config-llama2}]
{
    "mode": "eager",
    "dtype_in": "bf16",
    "hidden_size": 4096,
    "vocab_size": 32000,
    "intermediate_size": 11008,
    "actfn_algo": "pwl",
    "actfn_table_size": 256,
    "dtype_wts": "int4",
    "gemm_quant_scheme": "pergrp",
    "gemm_grpsize": 128,
    "bias": false,
    "rope_table_size": 4096,
    "num_heads": 32,
    "num_kv_heads": 32,
    "num_decoder_layers": 32,
    "kv_qscheme": "none",
    "max_position_embeddings": 4096,
    "mla": true,
    "q_lora_rank": 128,
    "kv_lora_rank": 128,
    "qk_nope_head_dim": 128,
    "qk_rope_head_dim": 64,
    "v_head_dim": 128,
}
    
\end{lstlisting}

\end{document}